\documentclass[preprint,authoryear,12pt]{elsarticle}

\usepackage{amssymb}
\usepackage{amsmath}
\usepackage{amsthm}
\usepackage{epstopdf}
\usepackage{setspace}
\usepackage{rotating}
\usepackage[T1]{fontenc}
\usepackage{subfig}
\usepackage{algorithm2e}

\usepackage{epstopdf}
\usepackage{adjustbox}
\usepackage{blindtext}
\journal{arXiv.org}
\begin{document}
\begin{frontmatter}
\title{A Novel Method for Comparative Analysis of DNA Sequences by Ramanujan-Fourier Transform}
\author{Changchuan Yin$^{1,\ast}$, Xuemeng E. Yin$^{2}$, Jiasong Wang$^{3}$}
\address{1.College of Information Systems and Technology, University of Phoenix, Chicago, IL 60601, USA\\
2.William Fremd High School, Palatine, IL 60067, USA\\
3.Department of Mathematics, Nanjing University, Nanjing, Jiangsu 210093, China\\
$\ast$ Corresponding author, Email: cyinbox@email.phoenix.edu
}
\begin{abstract}
Alignment-free sequence analysis approaches provide important alternatives over multiple sequence alignment (MSA) in biological sequence analysis because alignment-free approaches have low computation complexity and are not dependent on high level of sequence identity, however,  most of the existing alignment-free methods do not employ true full information content of sequences and thus can not accurately reveal similarities and differences among DNA sequences. We present a novel alignment-free computational method for sequence analysis based on Ramanujan-Fourier transform (RFT), in which complete information of DNA sequences is retained. We represent DNA sequences as four binary indicator sequences and apply RFT on the indicator sequences to convert them into frequency domain. The Euclidean distance of the complete RFT coefficients of DNA sequences are used as similarity measure. To address the different lengths in Euclidean space of RFT coefficients, we pad zeros to short DNA binary sequences so that the binary sequences equal the longest length in the comparison sequence data. Thus, the DNA sequences are compared in the same dimensional frequency space without information loss. We demonstrate the usefulness of the proposed method by presenting experimental results on hierarchical clustering of genes and genomes. The proposed method opens a new channel to biological sequence analysis, classification, and structural module identification.
\end{abstract}
\begin{keyword}
  Ramanujan-Fourier transform \sep  Similarity measure \sep Phylogenetic trees \sep Cluster analysis \sep DNA sequence
\end{keyword}
\end{frontmatter}
\section{Introduction}
\label{intro}
Comparative analysis of DNA sequences is a fundamental task in genome research. Multiple sequence alignment (MSA) has assumed a key role in comparative structure and function analysis of biological sequences, but MSA has high computation complexity and is dependent on high level of sequence identify. Several alignment-free sequence comparison methods have been proposed to address the problems limitations of MSA. These alignment-free methods can be divided into categories that include feature frequency profiles (FFP) \citep{sims2009alignment}, information theories \citep{li2001information}, graphic theory \citep{qi2011novel}, and analysis in frequency domain \citep{zhao2011novel}. In these methods, feature extraction or transformation of a sequence may incur loss of information and may not reflect the actual similarities and differences among DNA sequences. It has been one of major challenges in DNA sequence analysis to create accurate and efficient alignment-free methods for proper measurements of sequence similarity \citep{vinga2003alignment}. To this end, we explore a digital signal processing approach that may retain full information content to compare DNA sequences.

Ramanujan-Fourier transform (RFT), a time series transform method, is rediscovered in recent years and receives much attention. Ramanujan sums (RS) are named after Indian mathematician Srinivasa Ramanujan, who in 1918 introduced them and proved Vinogradov's theorem that every sufficiently large odd number is the sum of three primes \citep{ramanujan1918certain}. The applications of RS has expanded from initial number theory to time signal analysis by RFT \citep{gadiyar1999ramanujan}, including low-frequency noise processing \citep{planat2009ramanujan,planat2002ramanujan}, processing the shear component of the wind as Doppler spectrum estimation\citep{lagha2009doppler}, T-wave alternans analysis \citep{mainardi2008analysis}, sparse signal analysis \citep{chen2013sparse}, time frequency analysis \citep{sugavaneswaran2012time}, and protein structure classification \citep{mainardi2007}. These studies provided new insights and understanding special properties in RFT and opened new door in applying this method in different research domains.

Similar to the Fourier transform (FT), the RFT is orthogonal in nature and offers excellent energy conservation capacity \citep{chen2013matrix}. The RFT is operated on integers and hence can obtain a reduced quantization error implementation. Because of these excellent properties, we present a new similarity measure for DNA sequences using full information content derived from RFT of the DNA sequences. The results of applying the method on similarity analysis and hierarchical clustering of DNA sequences demonstrate that the RFT-based method is accurate and effective in DNA sequence comparison.

\section{Methods and Algorithms}
\label{sec:1}
\subsection{Representations of DNA sequences by binary indicators}
DNA molecules consist of four linearly linked nucleotides: adenine (A), thymine (T), cytosine (C), and guanine (G). A DNA sequence can be represented as a permutation of four characters $A, T, C$, and  $G$ at different lengths. Before signal processing methods are applied to symbolic DNA sequences, the sequences are converted to numerical series or binary indicator sequences \citep{Voss1992}. A DNA sequence denoted as, $x(1), x(2), \ldots, x(N)$, can be decomposed into four binary indicator sequences, $u_A(n), u_T(n), u_C(n)$, and $u_G(n)$ which indicate the presence or absence of four nucleotides, $A, T, C$, and $G$ at the $n-$th position, respectively. The indicator mapping of DNA sequences is defined as follows:
\begin{equation}
u_\alpha  (n) = \left\{ \begin{gathered}
  1,s(n) = \alpha  \hfill \\
  0,otherwise \hfill \\
\end{gathered}  \right.
\end{equation}
Where  $\alpha  \in \left\{ {A,T,C,G} \right\}, n=1, 2, ..., N $. The four indicator sequences correspond to the appearance of the four nucleotides at each position of the DNA sequence. For example, the indicator sequence, $u_A(n) =0001010111\ldots$, indicates that the nucleotide $A$ presents in the positions of 4, 6, 8, 9, and 10 of the DNA sequence.

\subsection{Ramanujan-Fourier transform}
In signal processing, transform methods are used to convert signals from time space into frequency space in order to estimate and analyze the informational content of the signal from different perspectives. Discrete Fourier Transform (DFT) is widely applied to periodic or quasi-periodic signals, but this technique is not appropriate for the analysis of aperiodic random signals. As a result, numerous methods have been developed to analyze aperiodic time series, Ramanujan-Fourier transform is recently rediscovered as a vital alternative to the Fourier transform \citep{carmichael1932expansions} with advantages of effectively processing time series with spectrums abundant with low frequencies \citep{planat2002ramanujan,planat2009ramanujan,planat20011}.

The Ramanujan sums $c_q (n)$ are the real sums of the $n-th$ power of the $q-th$ primitive roots of the unity \citep{ramanujan1918certain},
\begin{equation}
c_q (n) = \sum\limits_{p = 1,(p,q) = 1}^q {\exp (2i\pi \frac{p}
{q}n)}
\end{equation}
where $(p,q)=1$ indicates that p and q are relatively co-prime. The sums were introduced by Ramanujan as base functions over which typical arithmetical sequence or the original signal $x(n)$ may be projected as:
\begin{equation}
x(n) = \sum\limits_{q = 1}^\infty  {x_q c_q (n)}
\end{equation}
These base functions satisfy many suitable properties for signal decomposition such as multiplicative and orthogonality properties. Alternatively, the Ramanujan sums can be evaluated using the Euler totient function $\phi (q)$ and Moebius function $\mu (n)$ as follows \citep{lagha2009doppler}
\begin{equation}
c_q (n) = \mu \left( {\frac{q}
{{(q,n)}}} \right)\frac{{\phi (q)}}
{{\phi \left( {\frac{q}
{{(q,n)}}} \right)}}
\end{equation}
where the symbol $(q, n)$ denotes the great common factor of q and n. The equality (q, n) = 1 imposes q and n to be relatively co-prime. The Euler's totient function $\phi (q)$ is a multiplicative arithmetic function that counts the number of positive integers in the range $0 < k < n$ that are co-prime to n. The Euler's totient function $\phi (q)$ is defined as \citep{sandor2004handbook}
\begin{equation}
\phi (q) = q\prod\limits_i {\left( {1 - \frac{1}
{{q_i }}} \right)}
\end{equation}

The Moebius function $\mu (n)$ is also a multiplicative function and returns zero for positive integers and has its values in ${-1, 0, 1}$ depending on the factorization of n into prime factors. The Moebius function $\mu (n)$, is defined as \citep{sandor2004handbook}
\begin{equation}
\mu (n) = \left\{ \begin{gathered}
  0{\kern 1pt} {\kern 1pt} {\kern 1pt} {\kern 1pt} {\kern 1pt} {\kern 1pt} {\kern 1pt} {\kern 1pt} {\kern 1pt} {\kern 1pt} {\kern 1pt} {\kern 1pt} {\kern 1pt} {\kern 1pt} {\kern 1pt} {\kern 1pt} {\kern 1pt} {\kern 1pt} {\kern 1pt} {\kern 1pt} {\kern 1pt} {\kern 1pt} {\text{if}}{\kern 1pt} {\kern 1pt} {\text{n}}{\kern 1pt} {\kern 1pt} {\kern 1pt} {\text{content}}{\kern 1pt} {\kern 1pt} {\kern 1pt} {\text{squared}}{\kern 1pt} {\kern 1pt} {\text{value}}{\kern 1pt} {\kern 1pt} \beta _k  > 1{\kern 1pt} {\kern 1pt}  \hfill \\
  1{\kern 1pt} {\kern 1pt} {\kern 1pt} {\kern 1pt} {\kern 1pt} {\kern 1pt} {\kern 1pt} {\kern 1pt} {\kern 1pt} {\kern 1pt} {\kern 1pt} {\kern 1pt} {\kern 1pt} {\kern 1pt} {\kern 1pt} {\kern 1pt} {\kern 1pt} {\kern 1pt} {\kern 1pt} {\kern 1pt} {\kern 1pt} {\kern 1pt} {\kern 1pt} {\kern 1pt} {\text{if}}{\kern 1pt} {\kern 1pt} n = 1 \hfill \\
  ( - 1)^k {\kern 1pt} {\kern 1pt} {\kern 1pt} {\text{if}}{\kern 1pt} {\kern 1pt} {\text{n}}{\kern 1pt} {\kern 1pt} {\kern 1pt} {\text{is}}{\kern 1pt} {\kern 1pt} {\kern 1pt} {\text{the}}{\kern 1pt} {\kern 1pt} {\kern 1pt} {\text{product}}{\kern 1pt} {\kern 1pt} {\kern 1pt} {\text{of}}{\kern 1pt} {\kern 1pt} {\text{k}}{\kern 1pt} {\text{primer}}{\kern 1pt} {\kern 1pt} {\text{number}} \hfill \\
\end{gathered}  \right.
\end{equation}

By the relationship defined in (3), Carmicheal introduced the Ramanujan-Fourier transform (RFT) represented as \citep{carmichael1932expansions}
\begin{equation}
x_q  = \frac{1}
{{\phi (q)}}A_v (x(n)c_q (n))
\end{equation}
where is $A_v(g)$ the mean value of the function  $g(n)=x(n)c_q (n)$ and is defined as
\begin{equation}
A_v (g) = \mathop {\lim }\limits_{N \to \infty } \frac{1}
{N}\sum\limits_{n = 1}^N {g(n)}
\end{equation}

The Ramanujan-Fourier transform (RFT) of a signal $x(n)$ is defined as
\begin{equation}
x_q  = \frac{1}
{{\phi (q)}}\mathop {\lim }\limits_{N \to \infty } \frac{1}
{N}\sum\limits_{n = 1}^N {x(n)} c_q (n)
\end{equation}

Because it is a very time consuming task to compute $RS$ base functions $c_q (n)$, we retrieved the RS basis functions from pre-computed object instead of computing them online to reduce computation time of RFT. Recently, Chen et al introduced the 1D forward and inverse RFT of a signal $X$ of length N by matrix multiplication \citep{chen2013matrix}. Let $R$ be the RS matrix of size N-by-N and  defined as
\begin{equation}
R(q,j) = \frac{1}
{{\phi (q)N}}c_q (\bmod (j - 1,q) + 1),{\kern 1pt} {\kern 1pt} {\kern 1pt} {\text{q,}}{\kern 1pt} {\text{j}} \in [1,N]{\text{ }}
\end{equation}
where mod() is the modular operation. The forward 1D RFT and the inverse RFT of the signal $X$ can be realized as follows
\begin{equation}
\begin{gathered}
  Y = RX \hfill \\
  X = R^{ - 1} Y \hfill \\
\end{gathered}
\end{equation}

For technical reference, the following data are the firt 10 terms of Fibonacci numbers and their corresponding RFT coefficients, respectively: Fibonacci numbers = [1     1     2     3     5     8    13    21    34    55]; RFT = [14.3000    3.3000   -0.5500   -4.0000    3.9250   -5.8500   -0.8667    1.8000    2.9000    5.4250]. Because the RFT transform gives excellent results that retain the energy and inverse of the input signals. We investigate the usage of RFT as a measure of DNA sequence or protein sequence similarity. To this end, we apply RFT to the binary indicator sequence  $u_A ,u_C ,u_G ,u_T$ of a DNA sequence to obtain $R_A, R_G, R_G, R_T$, respectively. The RFT coefficients (amplitudes) of the four binary sequences of length $N$ are calculated as
\begin{equation}
R_\alpha  (q) = \frac{1}
{{\phi (q)N}}\sum\limits_{q = 1}^N {\sum\limits_{n = 1}^N {u_\alpha  (n)c_q (n)} } ,\alpha  \in \{ A,T,C,G\}
\end{equation}

Because the RFT coefficients for real value signal are real positive or negative numbers, the sum of the absolute values of RFT coefficients of the four binary sequences of a DNA sequence $PS$ is considered a power spectrum of the RFT of DNA sequences. It is used as a signature metric of DNA sequence in this study and is defined as
\begin{equation}
PS(q) = \sum\limits_{\alpha  \in \{ A,T,C,G\} } {\left| {R_\alpha  (q)} \right|}
\end{equation}

To address the different lengths of RFT coefficients in Euclidean distances, we pad zeros to short DNA binary sequences so that the binary sequences equal to the longest sequence length in the comparison data so that the DNA sequences are compared in the same dimensional Euclidean frequency space. We exclude the 1st term in RFT figure plotting because the 1st term is just the mean of real value signal and impacts scaling of figures, but the 1st term  is included for computing RFT distance because it is needed for recovering original data from RFT coefficients.

\subsection{Discrete Fourier transform}
Discrete Fourier Transform (DFT) is the transformation of N observation data (time domain) to $N$ new values (frequency domain). DFT spectral analysis of DNA sequences may detect latent and hidden periodical signals in the original sequences\citep{Shumway2010}. It may discover approximate repeats that are difficult to detect by direct tandem repeat search. Let
$U_{A},U_{T},U_{C},$ and $U_{G}$ be
the DFT of the binary sequences $u_A, u_T, u_C$, and $u_G$, the DFT of the numerical series $u_\alpha, \alpha \in \left\{ {A,T,C,G} \right\}$ of length N is defined as\\
 \begin{equation}\label{e:WeakBase}
  U_\alpha(k)=\sum_{n=1}^{N}u_\alpha(n)e^{-i\frac{2\pi}{N}kn}
\end{equation}
where $i = \sqrt{-1}$. The DFT power spectrum of the signal $u_\alpha$ at the frequency k is defined as
\begin{equation}
 PS(k) = \sum\limits_{\alpha \in \{ A,T,C,G\} } {\left| {U_\alpha (k)} \right|} ^2 ,k = 0,1,2, \cdots ,N - 1
\end{equation}
where $U[k]$ is the $k$-th DFT coefficient.\\

Due to the symmetric property of the DFT spectrum of real number signals, all figures of the DFT spectrum in this paper only show the first half of the original figures. In addition, we exclude the 1st term in DFT figure plotting because the 1st term DFT is just the sum of time data and also impacts scaling of figures.

\subsection{Distance metrics}
A distance metric $d\left( {x ,y } \right)$ is a nonnegative function on the set of pairs $\left( {x,y } \right)$ of finite sequences over a fixed alphabet. For example, one of its uses is a measure of the evolutionary change from DNA sequence $x$ to $y$. The evolutionary changes are reversible, and the fewest number of evolutionary changes is from $x$ to $y$ directly. Therefore, a distance metric shall be reflective, symmetric and transitive \citep{waterman1976, kanas1991,otu2003new}. A metric space is a set $X$ together with a metric $d$ on it. For example, the set of real numbers with the function $d(x,y) = \left| {x - y} \right|$ is a metric space. A distance metric in the metric space satisfies the following properties.\\
(1) $d(x,y) \ge 0$ for all $x,y \in X$; moreover, $d(x ,y ) = 0$, if and only if $x  = y$. \\
(2) $d(x ,y ) = d(y ,x )$  for all $x,y  \in X$. \\
(3) the triangle inequality, \emph{i.e.}, $d(x ,y ) \le d(x ,z ) + d(z ,y )$ for all $x,y,z  \in X$

The most common distance measure for time series is the Euclidean distance, which is the optimal distance measure for estimation if signals corrupted by additive Gaussian noise\citep{agrawal1993efficient,Yu2011}. The Euclidean metric distance of two time series $x_1 , \ldots ,x_n $ and  $y_1 , \ldots ,y_n $  is defined as
$$
d\left( {\left( {x_1 , \ldots ,x_n } \right),\left( {y_1 , \ldots ,y_n } \right)} \right) = \sqrt {\sum\limits_{k = 1}^n {(x_k  - y_k )^2 } }
$$
The Euclidean distances of power spectral of RFT of different DNA sequences are measured and used as a measure of similarity for these DNA sequences. The pairwise Euclidean distances of power spectral of RFT of DNA sequences are used to generate a similarity matrix, which can be used to construct a phylogenetic tree of these sequences. The phylogenetic trees constructed from a similarity matrix reflect classes information, hierarchical similarity and evolutionary relationships of the DNA sequences.

\subsection{Algorithm to compute pairwise distance of DNA sequences by RFT}
From above definitions and theories, we propose the following algorithm to calculate pairwise Euclidean distances as distance measures of DNA sequences by RFT, which is used  to construct similarity matrices for phylogenetic trees.

\begin{algorithm}[H]
 \SetAlgoLined
 \KwData{DNA SEQ1(length N1), SEQ2(length N2), SEQ3(length M), with $M>N1$ and $M>N2$}
 \KwResult{Pairwise distance of SEQ1, SEQ2 and SEQ3}
  Steps
 \begin{enumerate}
  \item Convert each SEQ1, SEQ2, SEQ3 to fours binary indicator sequences BS1, BS2, and BS3.
  \item Pad zeros to BS1 and BS2 to extend their lengths to M and become BS1M, BS2M, respectively.
  \item Compute the RFT coefficients of BS1M, BS2M, and BS3.
  \item Add the absolute values of RFT coefficients of four indicator sequences of each sequence, the resulted sum are RS1, RS2 and RS3.
  \item Compute the Euclidean distance d(RS1,RS2), d(RS2,RS3), and d(RS1,RS3) in M dimensional space.
\end{enumerate}
 \caption{Algorithm for calculating pairwise Euclidean distances of DNA sequences by RFT}
\end{algorithm}

\section{Results and Discussion}
\subsection{Ramanujan-Fourier transform on DNA sequences}
To illustrate the usage of RFT for the identification of hidden periodicities in signals, we applied RFT to the following periodic signal made up of sine and cosine signal with periodicties 10 and 20, and corrupted by white random noise:
\[
\begin{gathered}
  s[t] = \sin (2\pi \frac{t}
{{10}} + \frac{\pi }
{4}) + \cos (2\pi \frac{t}
{{20}} + \frac{\pi }
{4}) + noise \hfill \\
  t = 1:100 \hfill \\
\end{gathered}
\]

Figure 1(a) is the plot of the original periodic sine and cosine signal and shows that two periodicities 10 and 20 in the original sine and cosine functions are hidden by random noise. After RFT transforms, the two periodicities can be clearly identified as shown in Figure 1(b). Two pronounced peaks of RFT spectrum in Figure 1(b) at positions of $q=10$ and $q=20$ represent the two periodicities periodicities 10 and 20, respectively. For comparison, Figure 1(c) is the DFT spectrum plot of the signal. The two pronounced peaks of DFT spectrum in Figure 1(c) are at frequency $f=N/10, N=100$ and $f=N/20$ for $N=100$, respectively. From the illustrative example, we can see both DFT and RFT can reveal hidden periodicities. Furthermore, it is worth to  mention here is that a special property of RFT, which DFT does not hold, is that RFT may recover hidden periodicities in a phase modulated period signal \citep{planat2009ramanujan}.
\begin{figure}[tbp]
	\centering
	\subfloat[Original signal]{\includegraphics[width=3.0in]{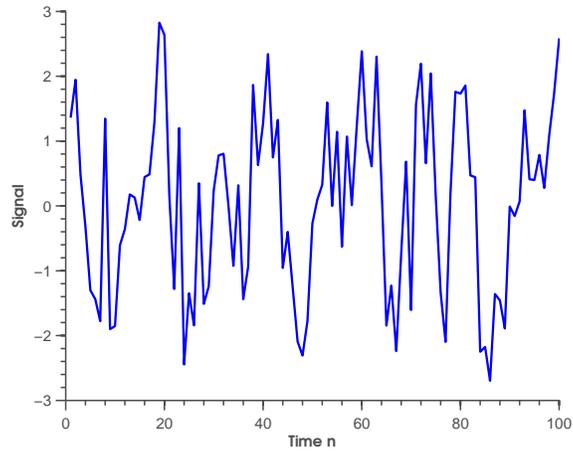}}\quad
	\subfloat[RFT transform]{\includegraphics[width=3.0in]{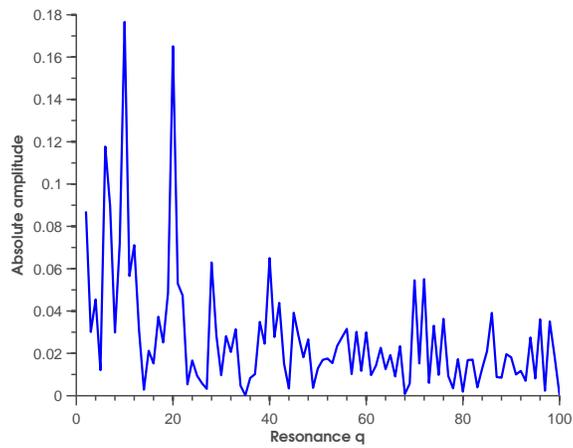}}\quad
    \subfloat[DFT transform]{\includegraphics[width=3.0in]{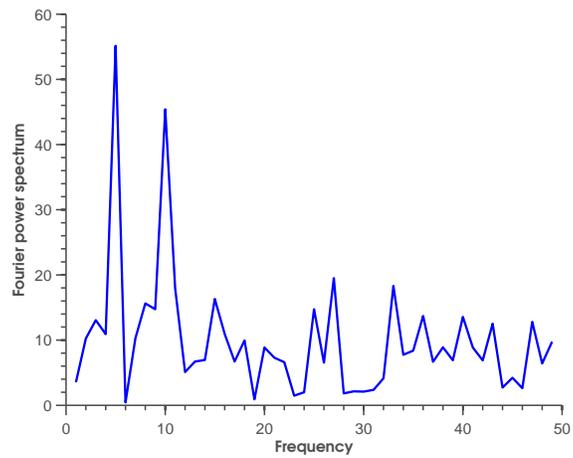}}
    \caption{Periodic signal collapsed with noise and its RFT and DFT transforms}
\end{figure}

To compare RFT with DFT in DNA sequence analysis, we applied RFT to spectrum analysis of exon and intron sequences. It is widely known that the 3-base periodicity, measured as the Fourier power spectrum of a DNA sequence at the frequency $N/3$, is present in most of exon sequences but not in the majority of intron sequences. This property has been used in gene finding algorithms \citep{tiwari1997prediction,jiang2008coding}. Previously we identified the origin of applied 3-base periodicity in DFT and applied it in protein coding predication in DNA sequences \citep{yin2005fourier,yin2006tracking, yin2007prediction,yin2008numerical} Figure 2(a) and 2(b) are the RFT and DFT spectra of the exon sequence of the Homo sapiens (human) mitochondrial cytochrome oxidase subunit I (COI) gene (Genbank ID = KC750830, N = 386 bp), respectively. As shown on the figure Figure 2(a) and (b), the exon sequence shows 3-base periodicity in DFT spectrum. The peak position in  the DFT spectrum is at frequency of $N/3$. The exon demonstrates a pronounced peak on $q=3$ position in the RFT spectrum. These results show that RFT can identify special 3-base periodicity in exons just as DFT can.
\begin{figure}[tbp]
	\centering
	\subfloat[RFT of exon]{\includegraphics[width=3.5in]{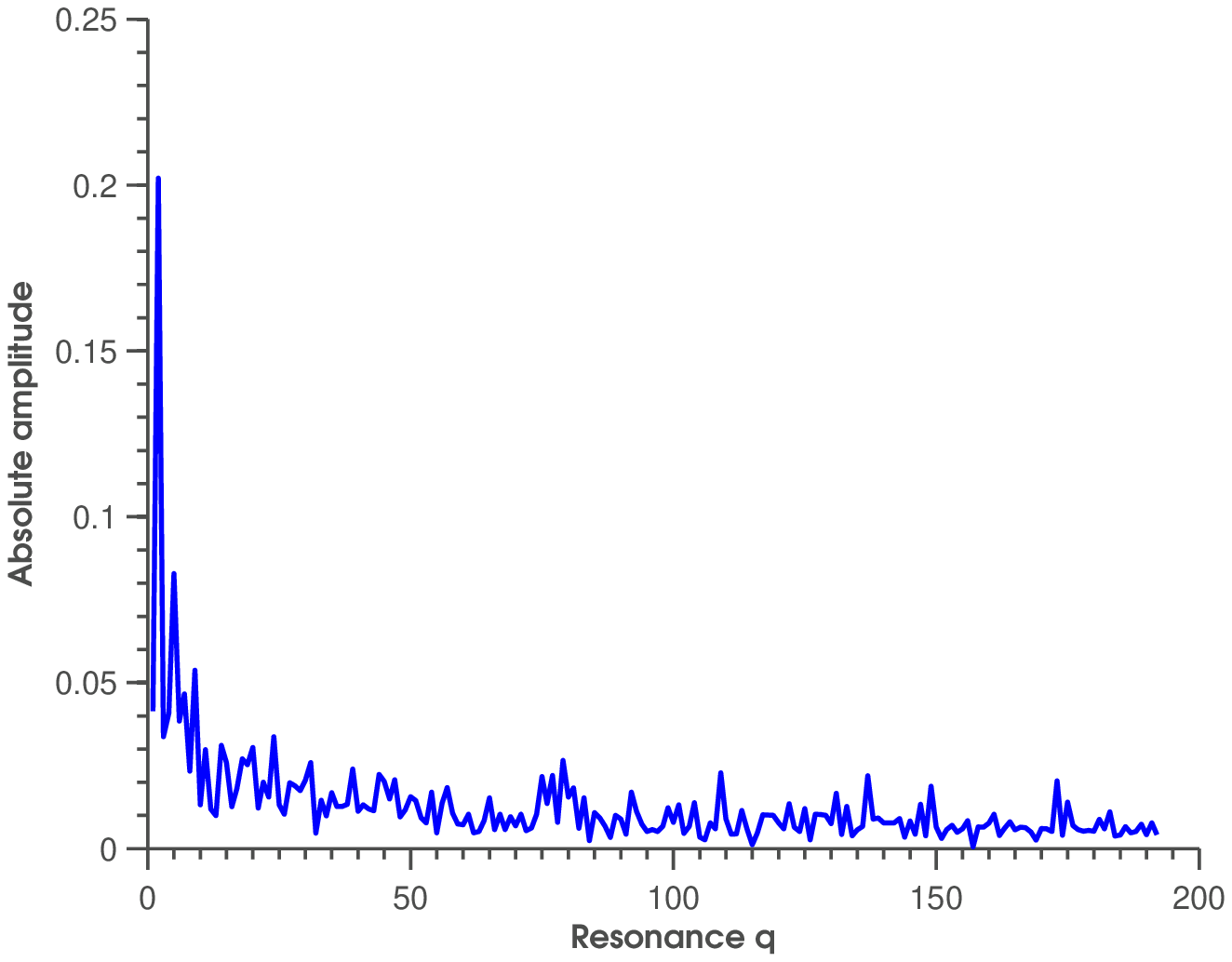}}\quad
	\subfloat[DFT of exon]{\includegraphics[width=3.5in]{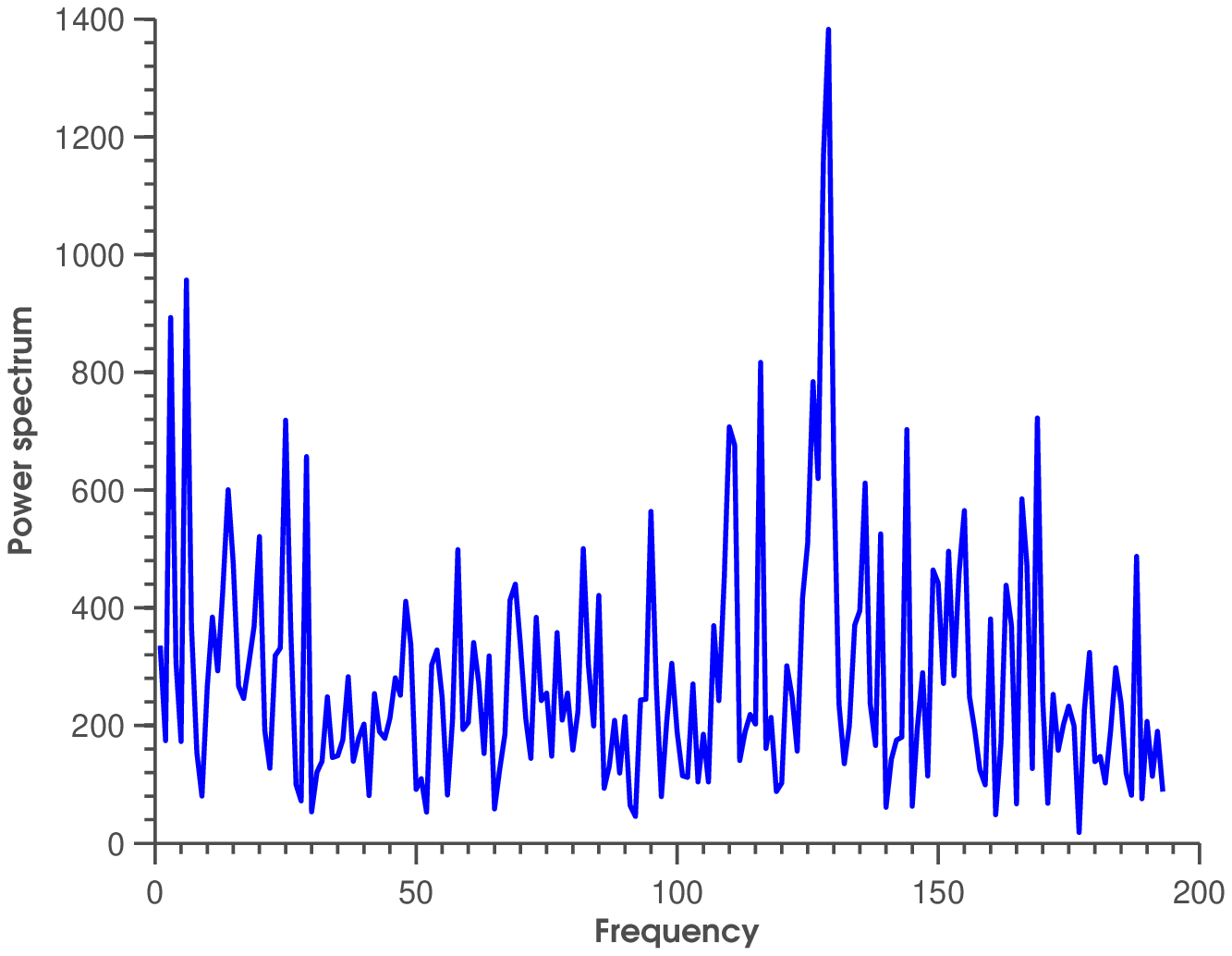}}
    \caption{RFT and DFT spectrum of exon sequence of Homo sapiens mitochondrial cytochrome oxidase subunit I (COI) gene (Genbank ID: KC750830, N = 386bp).}
\end{figure}

Figure 3(a) and 3(b) are the RFT and DFT spectra of the intron sequence of the Heteractis crispa isolate Hc86-02 cytochrome oxidase subunit I (COI) gene (GenBank ID: JQ918751, N=654 bp), respectively. For the purpose of comparison, the RFT plot of intron in Figure 3(a) was drawn in the same scale as the DFT exon plot in Figure 2(a). As shown in Figure 3(a) and 3(b), the intron does not shows 3-base periodicity in both DFT and RFT spectrum (q=3). The results indicates that RFT spectrum analysis is in agreement with DFT analysis of both exon and intron sequences.
\begin{figure}[tbp]
	\centering
	\subfloat[RFT of intron]{\includegraphics[width=3.5in]{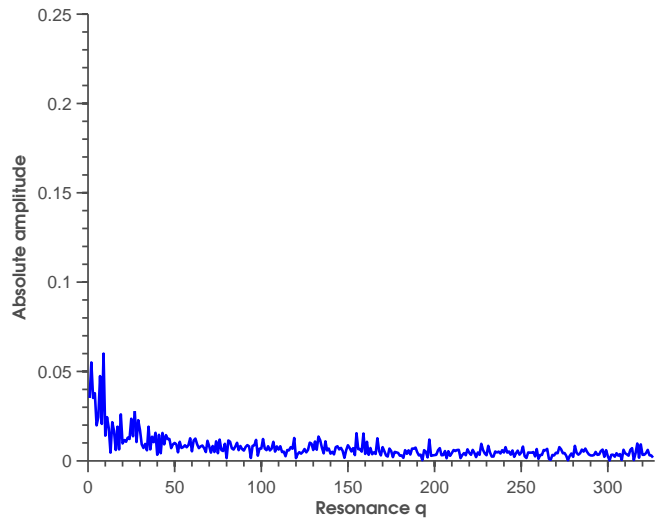}}\quad
	\subfloat[DFT of intron]{\includegraphics[width=3.5in]{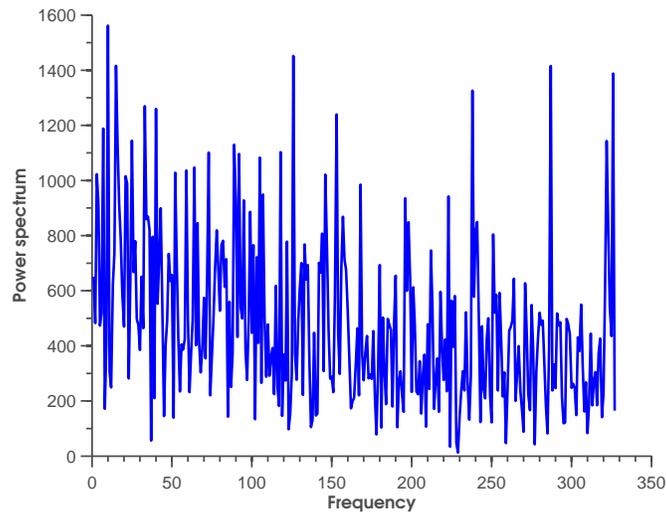}}
    \caption{RFT and DFT spectrum of intron sequence of Heteractis crispa isolate Hc86-02 cytochrome oxidase subunit I (COI) gene (GenBank ID = JQ918751, N = 654bp)}
\end{figure}
\subsection{Similarity metric of DNA sequences by RFT}
A common similarity measure between two DNA sequences is edit distance, which is defined as the minimum number of  substitutions (point mutations), insertions, deletions or genomic rearrangement needed to transform one sequence into the other during evolutionary process. The edit distance can be obtained by an optimal alignment of DNA sequences. Because we use all the RFT coefficients in frequency domain and the RFT coefficients have full information contents of the time domain from equation $11$. Thus the similarity measure by RFT in frequency domain is expected to correlate well with the edit distance rendered by point mutations and deletion mutations in DNA sequences.

To assess this correlation between the RFT distance measure and edit distance, we studied the relationship between original DNA sequences and a series of point mutations of the sequences using the RFT similarity metric. An intron sequence was introduced different numbers of point mutations at random positions in the sequence. The RFT distances between the mutants and the original sequence were measured. We tested the correlation of the RFT distances and number of point mutations. Figure 4 is the correlation between the amount of point mutations and the distance between the corresponding  mutations and original sequence. The result shows a linear relationship of RFT distances and the amount of point mutations. This result demonstrates the accuracy of the RFT distance metric on the difference of nucleotide mutations on the same length DNA sequences.
\begin{figure}
\begin{center}
 \includegraphics[width=4.00in]{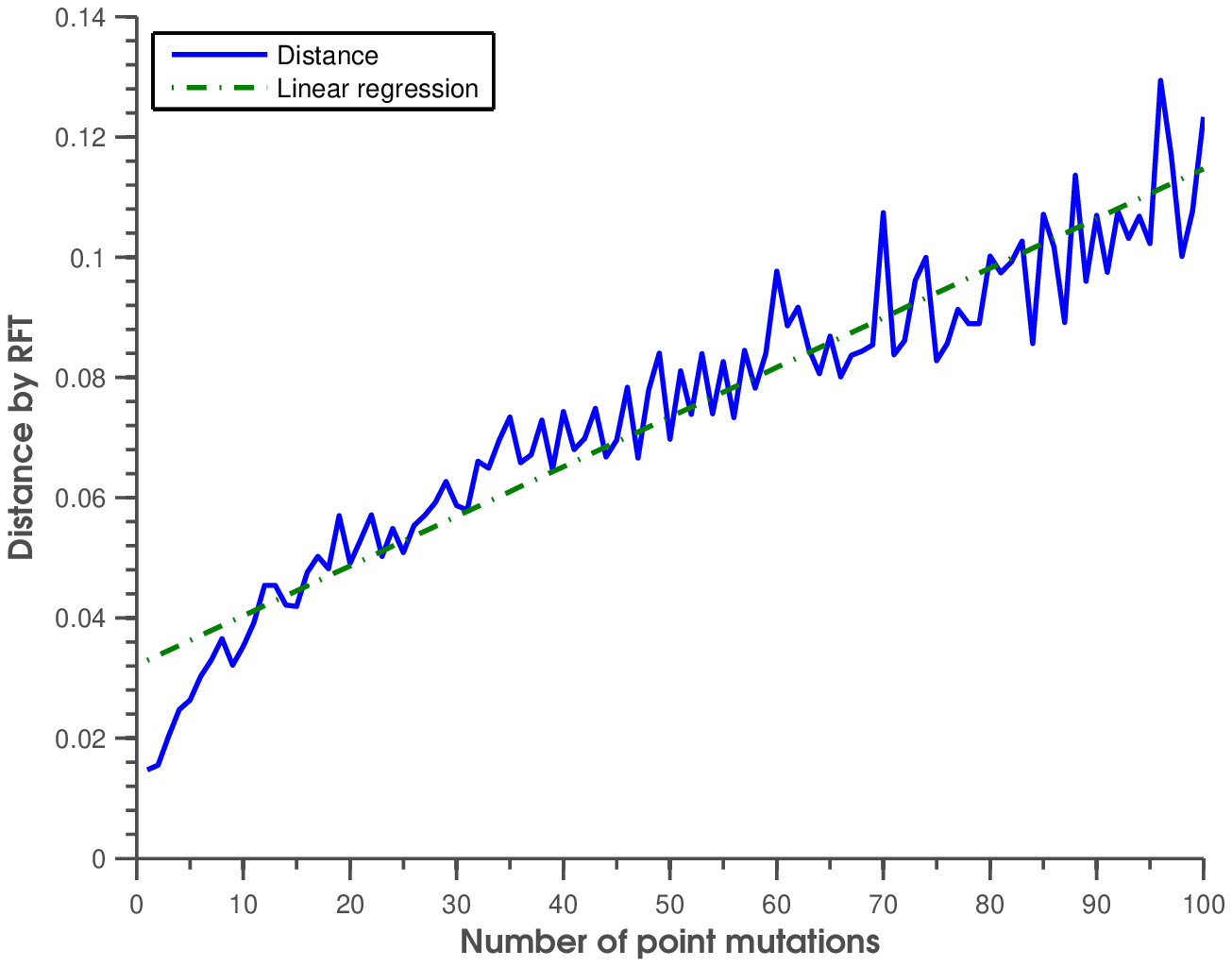}
\end{center}
\begin{small}
  \caption{Correlation between Ramanujan-Fourier transform distance and the number of point mutations of DNA sequences.}
\end{small}
\end{figure}
For an effective mathematical descriptors for similarity analysis, various mutation phenomena including deletions and insertions shall be considered simultaneously. Sequence alignment becomes unreliable or impossible for these diverse mutations. We assessed the accuracy of the proposed similarity distance metric using a series deletion mutations of a DNA sequence and measure the correlation of the similarity distance and deletion sizes in these mutants. An intron sequence is partially deleted from 3' end to generate different artificial mutants. The deletion size is from 1 bp to 100 bp. Then we measured the sequence distance between the mutants and the original sequence by the proposed RFT method. Figure 5 presents the correlation of the deletion lengths and the RFT distances between the corresponding deletion mutants and original sequence. The results show a sound linear relationship between the RFT distances and the deletion lengths of the mutants, thus indicating a robust and reliable behavior of the RFT distance metric in measuring similarities of sequences of different lengths. In this way, RFT can be used to compare rearrangements of sequences during evolutionary history.
\begin{figure}
\begin{center}
 \includegraphics[width=4.0in]{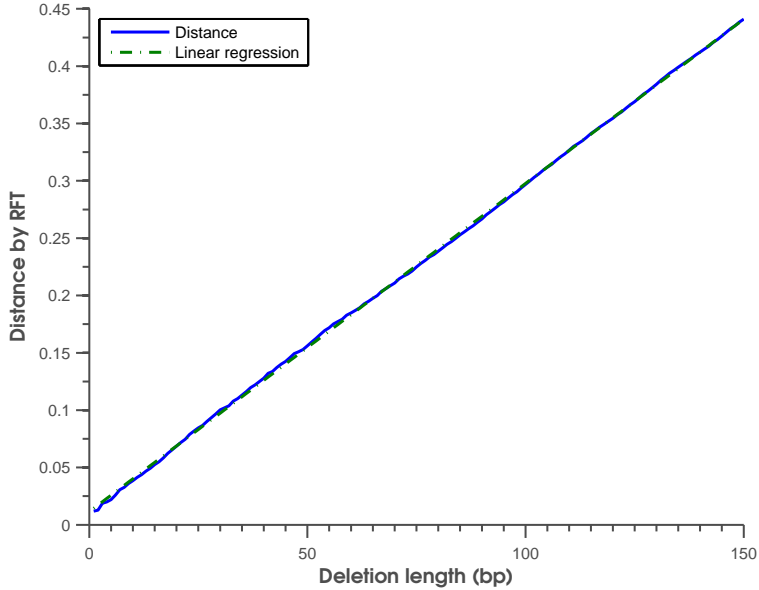}
\end{center}
\begin{small}
  \caption{Correlation of the RFT distance and the lengths of deletion mutants of DNA sequences.}
\end{small}
\end{figure}

To verify if the distance metric satisfies the triangle property, we randomly selected 200 exons from the Exon-Intron Database (EID) \citep{Shepelev2006} and measured pairwise distance of exons. For randomly chosen three exons as a test case, let $d1, d2$ and $d3$ be the three distance measures in RFT frequency domain and $d3$ be the largest distance. We compared the value of $d3$ and $d1+d2$ to validate the triangle property. Figure 6 is plot of the value $d3-(d1+d2)$ for triangle property test cases. The results shows all the positive values, showing that the RFT-based distance measure satisfies the triangle property and is thus a valid distance measure.
\begin{figure}
\begin{center}
 \includegraphics[width=4.00in]{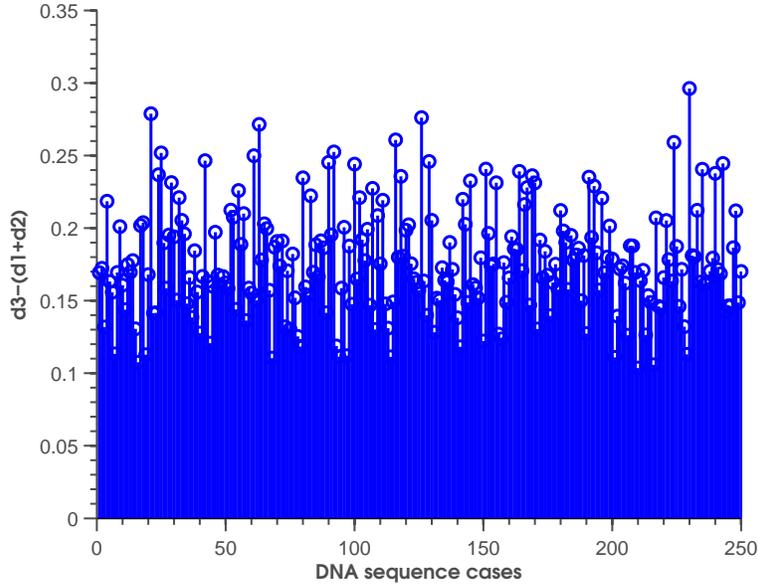}
\end{center}
\begin{small}
  \caption{Triangle property test of the RFT distances of DNA sequences.}
\end{small}
\end{figure}

\subsection{Application of the RFT distance metric in construction of phylogenetic trees}
To verify whether the similarity distance can be used for hierarchical clustering DNA sequences, we first generated mutations in DNA sequences and constructed phylogenetic trees from the pairwise RFT  distances of these mutants. We used an intron sequence as base and generated two new sequences A and B from the intron sequence using point mutations. 10\% of mutations were introduced into A and B.We then similarly evolved A into A1 and A2 and B into B1 and B2, using point mutations of 10\% of the sequences. We used the sequences A, A1, A2, B, B1 and B2 to build phylogenetic trees from similaritt matrix from the proposed RFT distance. The results in Figure 7 demonstrate that the proposed distance measure can be effectively used in construction of phylogenetic trees.

\begin{figure}
\begin{center}
 \includegraphics[width=4.0in]{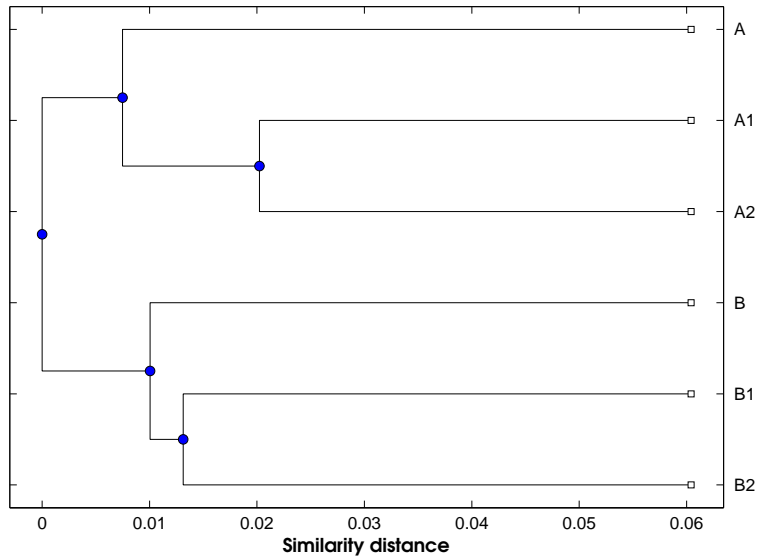}
\end{center}
\begin{small}
  \caption{Phylogenetic UPGMA tree of simulated DNA sequences using the RFT distances.}
\end{small}
\end{figure}

%
The effectiveness of the RFT measure in similarity analysis was assessed on the individual gene. Influenza A viruses can be divided into subtypes based on two proteins on the surface of the virus: hemagglutinin (HA) and neuraminidase (NA). The neuraminidase (NA) gene is associated with pandemic influenza and a wide range of natural hosts, including human, bird, and other animals.  The neuraminidase (NA) genes from different Influenza A virus subtypes were used in the test.  Figure 8 and Figure 9 are the phylogenetic trees of influzenza A virus constructed by the proposed RFT method and Jukes-Cantor method, respectively. Although both phylogenetic trees show correct grouping of different virus subtypes H7N9, H11N9, H3N2, and H1N1, the phylogenetic tree from RFT distance shows clearer branch differences than the tree from pairwise sequence alignment using the Jukes-Cartor distance. For example, the virus of highly homologous sequences such as A/Illinois H1N1 virus,06/2012,08/2012, and 01/2012,07/2012 cannot be separated by pairwise alignment using Jukes-Cartor distance, but they can be clearly separated with correct hierarchical relationship in the tree of RFT method. The other example in the figure is that the H7N9 virus mutants in China 2013 can only be clearly separated in the tree of RFT method. The hierarchical relationship among the H7N9 virus mutants in China is in agreement with the geographic distribution of the virus and the epidemiological investigation from previous findings \citep{xiong2013evolutionary}. Thus, the RFT distance tree can display clear levels of hierarchy and relationship among different virus, but alignment-based method cannot have clear spatial separation of similar species in the tree. We can depend on RFT tree analysis on the virus of one specie over different regions to see how the virus is evolved. These results demonstrate the superiority of the proposed RFT method on existing sequence alignment methods due to the fact the RFT distance is from calculation of all the sequence information and does not lose sequence information after transform. The resulted phylogenetic trees with RFT distance have the same topology, and are generally consistent with the reported results sequence alignment, and demonstrates a strong correlation with viral biology.
\begin{figure}
\begin{center}
 \includegraphics[width = 5.50in]{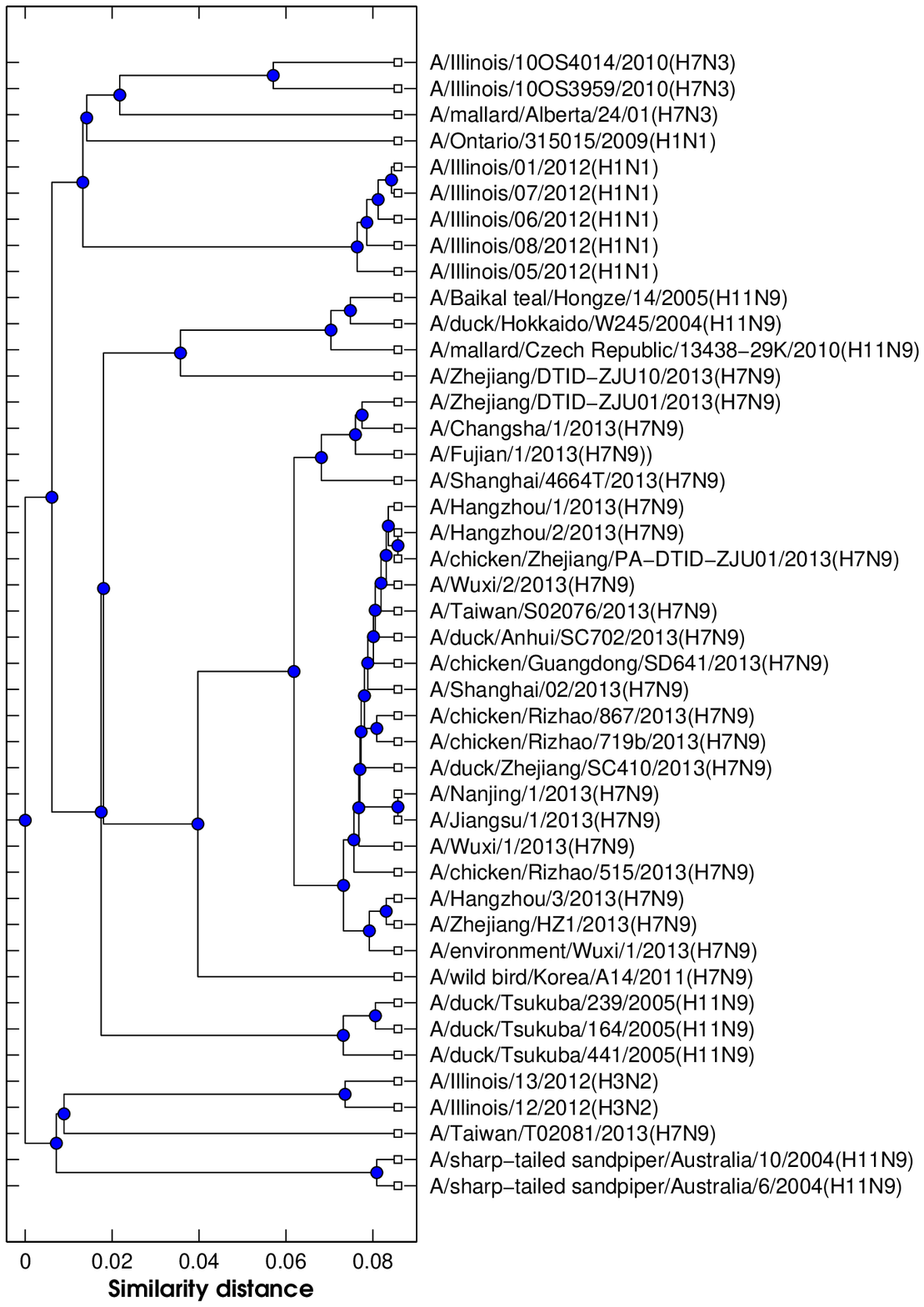}
\end{center}
\begin{small}
  \caption{Phylogenetic UPGMA tree of influenza A viruses using the RFT distances.}
\end{small}
\end{figure}
\begin{figure}
\begin{center}
 \includegraphics[width = 5.50in]{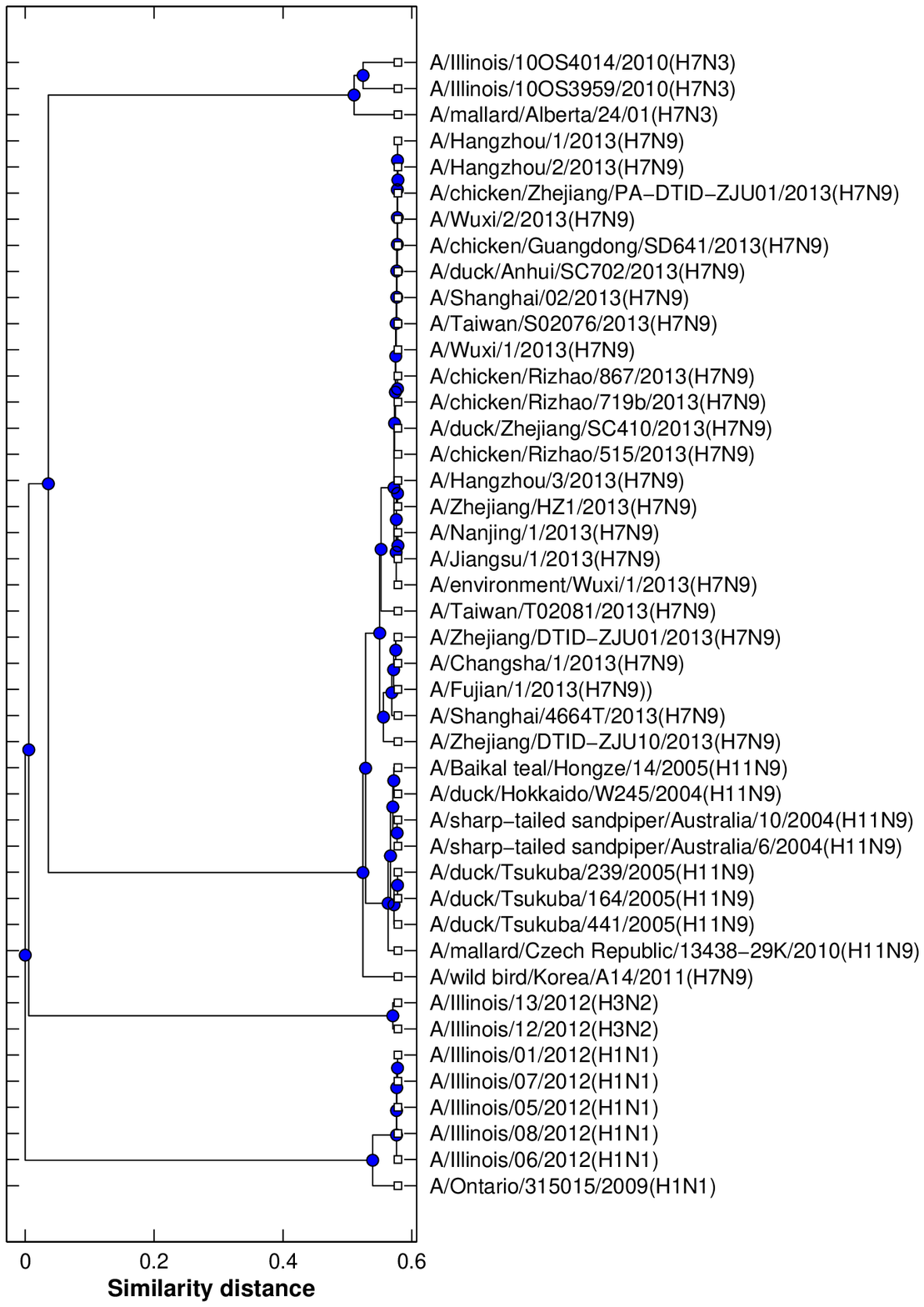}
\end{center}
\begin{small}
  \caption{Phylogenetic UPGMA tree of influenza A viruses using the sequence alignment Jukes-Cantor distance.}
\end{small}
\end{figure}

To compare the computational complexity of RFT measure on DNA analysis with DFT, we conducted a stress test by running 500 rounds DFT and 500 round RFT for a DNA sequence of 654 bp on a personal computer with 4 GB memory and a 2.4 GHz CPU. The total elapsed time of RFT test was 199.2686 seconds, while the corresponding elapse time for DFT was 20.4494 seconds. The result indicates RFT needs more computation time than DFT, but the performance was practicable for current computer power. From equation (9), RT shows $O(N^2 )$ arithmetical operations, while FFT has $O(N\operatorname{l} ogN)$ computational complexity. Another drawback of the RFT and DFT comparative analysis of DNA sequences is the limitation of the length of DNA sequences such as whole genomes. For very long DNA sequences or genomes, both RFT and DFT methods cannot handle the very long length due to limit of software memory. More studies are needed to address the problem for very long DNA sequences when using RFT as similarity mesaure. method.

One of the key tasks of the post-genome era is to determine the functional implications of gene or protein sequences. From similarity comparison and hierarchical clusterings, we may classify new sequences or genomes and infer their functions from classification. This requires accurate and efficient similarity measure for DNA sequences. The RFT based distance metric leads to accurate and reliable results in hierarchical clustering with accuracy. It shows a clear and better hierarchical tree than sequence alignments and sequence similarity scores for comparing amino acid and DNA sequences. Most alignment-free methods such as k-mer method and feature based methods may loss of information after extracting sequence or feature information. When using DFT power spectrum, the phase information in DFT is lost, but RFT coefficient real numbers and do not contain phase information, thus the RFT makes a reversible comprehensive map and characterization of a DNA sequence and thus retains all the sequence information for comparison.

\section{Conclusion}
In this work, we have established an effective similarity measure of DNA sequences based on RFT. We first performed RFT on DNA sequence after converting symbolic sequences to four binary indicator sequences. Euclidean distance is used to calculate the similarity of RFT coefficients. We conducted different DNA sequence mutants and assess the accuracy of the new RFT metric on the mutants. The similarity metrics have been evaluated by constructing phylogenetic trees of virus at gene levels. Our work provides encouraging steps for applying the rediscovered RFT approach for effective comparative studies on biological sequences.

\bibliographystyle{elsarticle-harv}
\bibliography{../References/myrefs}

\section*{Supplementary Materials}
The following tables are the GenBank access numbers and related information of the data used in the paper.
\begin{table}[ht]
\caption{Genbank accession numbers of influenza virus A used in the paper} 
\centering 
\begin{tabular}{l l} 
\hline\hline 
Genbank ID & Name \\ [0.5ex] 
\hline 
KC891137 & A/Illinois/05/2012(H1N1)\\
      KC891134 & A/Illinois/06/2012(H1N1)\\
      KC891128 & A/Illinois/08/2012(H1N1)\\
      KC891564 & A/Illinois/01/2012(H1N1)\\
      KC891131 & A/Illinois/07/2012(H1N1)\\
      KC893127 & A/Illinois/13/2012(H3N2)\\
      KC893131 & A/Illinois/12/2012(H3N2)\\
      DQ017515 & A/mallard/Alberta/24/01(H7N3)\\
      CY060664 & A/Ontario/315015/2009(H1N1)\\
      JF789604 & A/mallard/Czech Republic/13438-29K/2010(H11N9)\\
      GQ184333 & A/Baikal teal/Hongze/14/2005(H11N9)\\
      KF021599 & A/Shanghai/02/2013(H7N9)\\
      KC885958 & A/Zhejiang/DTID-ZJU01/2013(H7N9)\\
      KC994454 & A/Fujian/1/2013(H7N9)\\
      KC853765 & A/Hangzhou/1/2013(H7N9)\\
      KF001514 & A/Hangzhou/2/2013(H7N9)\\
      KF001517 & A/Hangzhou/3/2013(H7N9)\\
      KC896776 & A/Nanjing/1/2013(H7N9)\\
      KC853231 & A/Shanghai/4664T/2013(H7N9)\\
      KF018055 & A/Taiwan/T02081/2013(H7N9)\\
      KF018047 & A/Taiwan/S02076/2013(H7N9)\\
      CY147062 & A/duck/Anhui/SC702/2013(H7N9)\\
      CY147070 & A/duck/Zhejiang/SC410/2013(H7N9)\\
      CY146910 & A/chicken/Guangdong/SD641/2013(H7N9)\\
      JN244222 & A/wild bird/Korea/A14/2011(H7N9)\\
      CY029883 & A/sharp-tailed sandpiper/Australia/10/2004(H11N9)\\
      AB298284 & A/duck/Hokkaido/W245/2004(H11N9)\\
      AB472035 & A/duck/Tsukuba/239/2005(H11N9)\\
      CY025199 & A/sharp-tailed sandpiper/Australia/6/2004(H11N9)\\
      AB472034 & A/duck/Tsukuba/164/2005(H11N9)\\
      AB472033 & A/duck/Tsukuba/441/2005(H11N9)\\
\hline 
\end{tabular}
\label{table:nonlin} 
\end{table}






\end{document}